\documentclass[osajnl,twocolumn,showpacs,superscriptaddress,10pt]{revtex4-1} %% use 11pt for Applied Optics
\usepackage{amsmath,amssymb,graphicx}
\usepackage{xcolor}
\usepackage{soul}
\begin{document}

\title{Enhancement of the sensitivity of a temperature sensor based on Fiber Bragg Gratings via weak value amplification}

\author{L. J. Salazar-Serrano}\email{Corresponding author: luis-jose.salazar@icfo.es}
\affiliation{ICFO-Institut de Ciencies Fotoniques, Mediterranean
Technology Park, 08860 Castelldefels (Barcelona), Spain}
\affiliation{Quantum Optics Laboratory, Universidad de los Andes,
AA 4976, Bogot\'{a}, Colombia}

\author{D. Barrera}
\affiliation{iTEAM Institute, Universidad Politecnica de Valencia,
46022 Valencia, Spain}

\author{W. Amaya}
\affiliation{ICFO-Institut de Ciencies Fotoniques, Mediterranean
Technology Park, 08860 Castelldefels (Barcelona), Spain}

\author{S. Sales}
\affiliation{iTEAM Institute, Universidad Politecnica de Valencia,
46022 Valencia, Spain}

\author{V. Pruneri}
\affiliation{ICFO-Institut de Ciencies Fotoniques, Mediterranean
Technology Park, 08860 Castelldefels (Barcelona), Spain}

\author{J. Capmany}
\affiliation{iTEAM Institute, Universidad Politecnica de Valencia,
46022 Valencia, Spain}

\author{J. P. Torres}
\affiliation{ICFO-Institut de Ciencies Fotoniques, Mediterranean
Technology Park, 08860 Castelldefels (Barcelona), Spain}
\affiliation{Dep. of Signal Theory and Communications, Universitat
Politecnica de Catalunya, 08034 Barcelona, Spain}

\begin{abstract}
We present a {\em proof-of-concept} experiment aimed at increasing
the sensitivity of temperature sensors implemented with Fiber
Bragg gratings by making use of a {\em weak value amplification}
scheme. The technique requires only linear optics elements for its
implementation, and appears as a promising method for extending
the range of temperatures changes detectable to increasingly lower
values than state-of the-art sensors can currently provide. The
device implemented here is able to generate a shift of the
centroid of the spectrum of a pulse of $\sim 0.035$ nm/$^\circ$C,
a nearly fourfold increase in sensitivity over the same Fiber
Bragg Grating system interrogated using standard methods.

\end{abstract}

%140.3490 Fiber Bragg gratings;
%120.3180   Interferometry
%060.2370 Fiber optics sensors.
%120.6780   Temperature

\maketitle %% required
Fiber Bragg Gratings (FBG) constitute nowadays a key ingredient of
many devices used in communication and sensing applications
\cite{book_kashyap}. They can easily be integrated in all-fiber
systems, their dielectric nature make them non-conducting and
immune to electromagnetic interference, and current technology
allows to tailor the properties of FBGs to adapt to the specific
requirements of each application.

When considering a FBG as a sensor, it can be understood as a
bandpass filter whose central frequency depends on the value of a
variable (temperature or strain). Thus the achievement
of increasingly higher sensitivity in FBG-based systems implies
the development of new techniques to enhance the shift of the
central frequency of the filter for a given change of temperature or strain.

Here we consider the use of a technique generally referred as {\em
weak value amplification} (WVA), a concept first introduced by
Aharonov, Albert, and Vaidman \cite{aharonov1988}. It is a signal
enhancement ({\em amplification}) technique which is used in
metrology applications to measure tiny changes of a variable that
cannot be detected otherwise because of technical limitations,
i.e., the insufficient sensitivity of the detection system. It
makes use of the weak coupling that is introduced between two
degrees of freedom of a system. Here, the weak coupling will take
place between the shift of the centroid of the spectrum of a pulse
and its polarization.

The concept of weak value amplification can be readily understood
in terms of constructive and destructive interference between
probability amplitudes in a quantum mechanics context
\cite{duck1989}, or in terms of interference of classical waves
\cite{howell2010,high_signal2012}. Indeed, most of the
experimental implementations of the concept, since its first
demonstration in 1991 \cite{ritchie1991}, belong to the last type.
In this scenario, the usefulness of weak value amplification for
measuring extremely small quantities has been demonstrated under a
great variety of experimental conditions
\cite{hosten2008,ben_dixon2009,howell2010_freq,xu_guo2013,salazar2014}.

The use of FBGs as temperature sensors have been considered
\cite{tahir2009,ricchiuti2014}. Recently the WVA concept was
applied to the demonstration of a laser-based thermostat, based on
the measurement of the temperature-induced deviation of a laser
when traversing a fluid with a high thermo-optic coefficient
\cite{egan2012}. In this letter we show the usefulness of this
technique when applied to temperature sensing based on the use of
FBGs. It is important to notice that while we consider a
temperature sensor, other characteristics as well, such as
strain could have also be considered as targets.

The system considered makes use of a broad-band light source, two
FBGs at slightly different temperatures, and an optical spectrum
analyzer (OSA), all of them interconnected by optical circulators
and optical fibers. The FBG reflects only a small portion of the
input spectrum centered at a certain value determined by the
combination of the temperature and strain to which the device is
subjected. For configurations where the FBG is isolated from any
source of strain, the position of the centroid of the spectrum of
the reflected light varies linearly with respect to the
temperature with a sensitivity ranging from
$0.08\,\mathrm{nm/^{\circ}C}$ to $0.014\,\mathrm{nm/^{\circ}C}$
determined by the material of the fiber. As a result spectrum
analyzers with high resolution are required to measure
temperatures changes below one degree centigrade. Here we
demonstrate a system that can reach a sensitivity of up to $\sim
0.035\,\mathrm{nm/^{\circ}C}$ when WVA is used, to be compared
with a sensitivity of $\sim 0.009\,\mathrm{nm/^{\circ}C}$ that we
measure without the use of WVA. This enhancement allows to measure
smaller temperature difference given a specific sensitivity of the
OSA.

\begin{figure}[t!]
\centering
\includegraphics[width=0.4\textwidth]{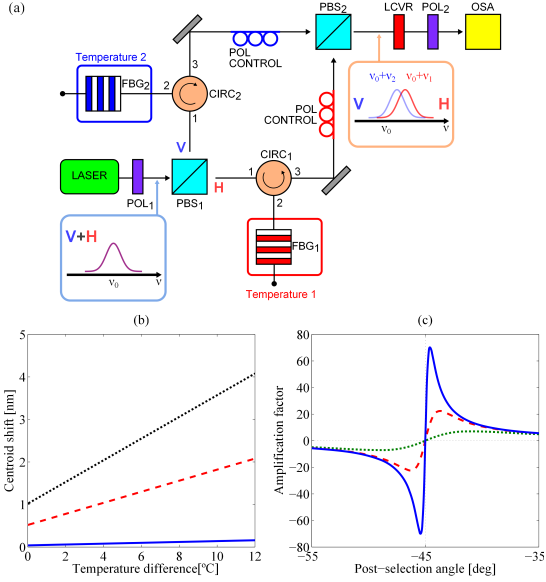}
\caption{(a) Experimental scheme. $\mathrm{FBG_1}$ and
$\mathrm{FBG_2}$: Fiber Bragg Gratings; $\mathrm{POL_1}$ and
$\mathrm{POL_2}$: polarizers; $\mathrm{PBS_1}$ and
$\mathrm{PBS_1}$: polarizing beam splitters; LCVP: Liquid-crystal
variable retarder; OSA: Optical Spectrum Analyzer;
$\mathrm{CIRC_1}$ and $\mathrm{CIRC_2}$: optical circulators. (b)
Theoretical shift of the centroid of the output spectrum as a
function of the temperature difference $T_1-T_2$ for three
different post-selection polarizations, and thus amplification
factors. $\beta=0^{\circ}$ and $\mathcal{A}=1$ (blue continuous
line), $\beta=-42.8^{\circ}$ and $\mathcal{A}=25$ (red dashed
line), and $\beta=-44.02^{\circ}$ and $\mathcal{A}=50$ (black dotted line).
(c) Theoretical amplification factor as a function
of the post-selection angle $\beta$ for different values of
$\gamma \cos \delta$: $0.99$ (continuous line), $0.999$
(dashed-line) and $0.9999$ (dotted-line).}\label{fig:figure1}
\end{figure}

Figure \ref{fig:figure1}(a) depicts the experimental scheme
implemented. A laser generates pulses with central frequency
$\nu_0$ (central wavelength: $1549\,\mathrm{nm}$) that are
linearly polarized at $+45^{\circ}$ by using a linear polarizer
($\mathrm{POL_1}$). {\em This constitute the pre-selection stage}.
The laser is a femtosecond fiber laser (Calmar Laser - Mendocino)
that generates $320\,\mathrm{fs}$ pulses (bandwidth:
$11\,\mathrm{nm}$) with a Gaussian-like spectrum, average power
$3\,\mathrm{mW}$ and repetition rate $20\,\mathrm{MHz}$.

The two orthogonal polarizations ($H$ and $V$) are divided by a
polarizing beam splitter ($\mathrm{PBS_1}$) and follow different
paths. The signal in each path is connected to FBGs
($\mathrm{FBG_1}$ and $\mathrm{FBG_2}$) by means of circulators
($\mathrm{CIRC_1}$ and $\mathrm{CIRC_2}$). The signal traversing
each arm of the interferometer is focused into a single-mode (SM)
fiber that is connected to the first port the circulator. The
second port is connected to the FBG that filters out the input
signal with an efficiency of $14\%$, and leaves the FBG with a
Gaussian-like spectrum of $2\,\mathrm{nm}$ centered at $\approx
1551\,\mathrm{nm}$.

Each FBG is embedded into an oven that is set to a different
temperature, $T_1$ and $T_2$. The FBG acts as a filter whose
central frequency is determined by the temperatures of the
corresponding oven. If the bandwidth of the input pulse is larger
than the bandwidth of the FBG, the effect of a temperature
difference is to generate two similar pulses with orthogonal
polarizations and different central frequencies. The third port of
each circulator is connected to a collimator lens that launches
the output beams towards a second PBS ($\mathrm{PBS_2}$), that
combines the two pulses into a single beam.

Before reaching ($\mathrm{PBS_2}$), and due to polarization
changes introduced by the circulators and FBGs, the state of
polarization of each pulse is rectified after leaving the
circulators for the second time using polarization controllers. In
this way, we assure that before recombining again the two pulses
in $\mathrm{PBS_2}$, the pulse that traversed $\mathrm{FBG_1}$ is
horizontally polarized and its central frequency is $\nu_1$,
whereas the pulse that traversed $\mathrm{FBG_2}$ is vertically
polarized and is spectrum is centered at $\nu_2$. In all cases, we
are interested in detecting small temperature changes, so that the
frequency shift $\nu_1-\nu_2$ is smaller than the FBG bandwidth
($B$). To compensate the phase introduced due to birefringence in
the circulators and  Single-mode (SM) fibers, a Liquid Crystal
Variable Retarder (LCVR) (Thorlabs - LCC1113-C) is added after
$\mathrm{PBS_2}$.

After $\mathrm{PBS_2}$ the electric field in the frequency domain
reads
\begin{eqnarray}
& & \textbf{E}(\nu)=\frac{E_0}{\sqrt{2}} \left\{ \hat{x} \exp\left[-\frac{(\nu-\nu_0-\nu_1)^2}{2B^2}\right] \right. \nonumber \\
& & + \left. \hat{y} \exp\left[-\frac{(\nu-\nu_0-\nu_2)^2}{2B^2}+i
(2\pi\nu\tau+\delta) \right] \right\}\label{eq:Eweak}
\end{eqnarray}
where $\hat{x}$ and $\hat{y}$ designate horizontal and vertical
polarization, respectively, $\tau$ takes into account the optical
path difference present in the experimental setup, and $B^2=\pi^2
T^2/\ln2$, with $T$ being the temporal duration (full width half
maximum) of the pulses reflected from the FBGs.
$\delta=\phi-\Gamma$, where $\phi$ denotes a phase due to the
birefringence induced from bends and twists in circulators and
single-mode fibers and $\Gamma$ is a phase introduced with a
Liquid Crystal Variable Retarder ($\mathrm{LCVR}$) to compensate
the unwanted phase $\phi$. Inspection of Eq. (\ref{eq:Eweak})
shows clearly the coupling between the shift of the centroid of
the spectrum of each pulse, $\nu_1$ and $\nu_2$, and its
polarization, a key element of the WVA scheme.

The weak value amplification effect is introduced by projecting
the recombined signal into a polarization state ${\bf
e}_{\mathrm{out}}=\cos\beta\hat{x}+\sin\beta\hat{y}$ with the help
of a second polarizer ($\mathrm{POL_2}$) that is rotated using a
motorized rotation stage. {\em This is the post-selection stage}.
After the post-selection, the output beam is collimated with a SM
fiber connected to an Optical Spectrum Analyzer (OSA).

The power spectrum ($S$) measured with the OSA, after post-selection, reads
\begin{eqnarray}
& & S (\nu)=\frac{S_0}{2} \left\{ \cos^2\beta\exp\left[-\frac{(\nu-\nu_0-\nu_1)^2}{B^2}\right] \right. \nonumber \\
& & +\sin^2\beta\exp\left[-\frac{(\nu-\nu_0-\nu_2)^2}{B^2}\right] \\
& & +\gamma\exp\left[-\frac{(\nu-\nu_+)^2}{B^2}\right]
\cos(2\pi\nu\tau+\delta)\nonumber \bigg\}
\label{eq:Sout}
\end{eqnarray}
where $\nu_{+}=(\nu_1+\nu_2)/2$, $\nu_{-}=(\nu_1-\nu_2)/2$ and
$\gamma=\exp(-\nu_{-}^2/B^2)$. After post-selection, the beams
reflected from each FBG interfere \cite{salazar_interference}. As
a result, there is a reshaping of the output spectrum.

\begin{figure}[t!]
\centering
\includegraphics[width=0.45\textwidth]{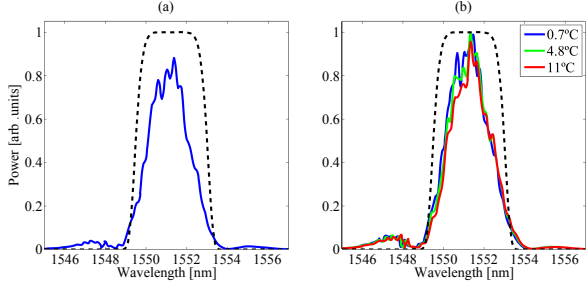}
\caption{(a) Spectrum of the signal reflected from
$\mathrm{FBG_2}$ at a fixed temperature ($\beta=-90^{\circ}$), and
(b) from $\mathrm{FBG_1}$ at different temperatures
($\beta=0^{\circ}$). Dashed lines: transmission function of the
super-Gaussian filter used to get rid of the unwanted side lobes
present in the signal.}\label{fig:figure2}
\end{figure}

Both FBG can show slightly different spectral responses due to
errors in the fabrication process. We keep one of the FBGs at a
constant temperature $T_2$, and measure its reflectivity spectrum
to be centered at $\nu_2^0$. The other FBG is used to measure a
variable temperature $T_1$, and the centroid of the spectrum of
the reflected signal is assumed to change linearly with
temperature as
\begin{equation}
\nu_1(T_1)=\nu_1^0+\kappa (T_1-T_2). \label{response_FBG}
\end{equation}
The centroid of the spectrum at the output port of the
interferometer is $\langle \nu\rangle=\int \nu S(\nu)d\nu / \int
S(\nu)d\nu$, where $S$ is measured with the $\mathrm{OSA}$. In the
weak coupling regime, the temporal delay $\tau$ in the
interferometer is much smaller than the pulse duration $T$, and
the frequency shifts $\nu_1(T_1)$ and $\nu_2^0$ are small compared
to the bandwidth $B$. In this scenario, the centroid of the
spectrum writes
\begin{equation}
\langle \nu\rangle=\nu_0+\nu_{+}+\mathcal{A}\,\nu_{-}\,.
\label{eq:DeltaNu1}
\end{equation}
where
\begin{equation}
\mathcal{A}=\frac{\cos2\beta}{1+\gamma \sin2\beta\,\cos \delta}
\label{amplification_factor}
\end{equation}
is the amplification factor.

\begin{figure}[t!]
\centering
\includegraphics[width=0.45\textwidth]{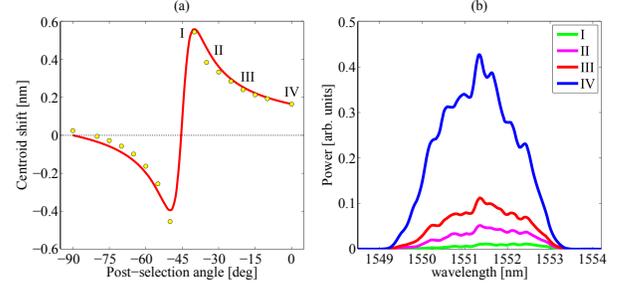}
\caption{(a) Shift of the centroid of the spectrum for
$T_1-T_2=11^{\circ}\mathrm{C}$ and different post-selection angles
(dots) in the interval $-90^{\circ}\leq\beta\leq0^{\circ}$. (b)
Output spectrum measured for some selected cases:
$\beta=-40^{\circ}$ (I), $\beta=-35^{\circ}$ (II),
$\beta=-25^{\circ}$ (III) and $\beta=0^{\circ}$
(IV).}\label{fig:figure3}
\end{figure}

We take as reference for the measurements $\nu_2^0$, which is
measured for an angle $\beta=-90^{\circ}$. One can easily find from
Eqs. (\ref{response_FBG}) and (\ref{eq:DeltaNu1}) that the shift
of the centroid $\langle \Delta \nu \rangle=\langle \nu
\rangle-\nu_2^0$ is
\begin{equation}
\langle \Delta \nu
\rangle=\frac{\kappa}{2}(\mathcal{A}+1)(T_1-T_2)+(\mathcal{A}+1)\left(\frac{\nu_1^0-\nu_2^0}{2}\right).
\label{deltanu_ref}
\end{equation}
The shift of the centroid of the spectrum is proportional to the
difference in temperature between the FBGs. If we project into a
polarization state that selects only the signal coming from the
FBG with a variable temperature ($\beta=0^{\circ}$), then
$\mathcal{A}=1$ and the constant of proportionality turns out to
be $\kappa$, which is determined by the response of the FBG to
changing temperatures. However, when we project into different
polarization states, $\kappa$ is multiplied by the amplification
factor $\mathcal{A}$ that can be much larger than one. Figure
\ref{fig:figure1} (b) shows the shift of the centroid of the
spectrum, expressed in terms of wavelength shift, for three values
of the amplification factor that corresponds to three different
output polarization projections. Notice the large enhancement of
wavelength shift that can be achieved for a given temperature
difference $T_1-T_2$ when different output polarization states are
selected.

The amplification factor can be very large. However, in practice,
its maximum value is limited by different experimental factors. On
the one hand, it strongly depends on how well the phase introduced
by the single-mode fibers $\phi$ is compensated by the variable
retarder (LCVR), as shown in Fig. \ref{fig:figure1} (c). From Eq.
(\ref{amplification_factor}), we obtain that the maximum
amplification factor that can be achieved for an uncompensated
phase is $\mathcal{A}_{\mathrm{max}}=(1-\cos^2 \delta)^{-1/2}$,
which is obtained for the post-selection angle
$-\pi/4^{\circ}\pm\beta_{\mathrm{max}}+\frac{1}{2}\arcsin(\cos
\delta)$. The largest enhancement is obtained when $\phi=\Gamma$.
On the other hand, in any weak value amplification scenario there
is attenuation of the amplitude of the output signal. For low
amplification factors, this attenuation can be similarly small.
However, a large amplification factor is accompanied by a large
attenuation, since the input and post-selected polarization are
nearly orthogonal. The amplification factor achievable is thus
limited by the Signal-to-Noise ratio available at the detection
stage.

\begin{figure}[t!]
\centering
\includegraphics[width=0.35\textwidth]{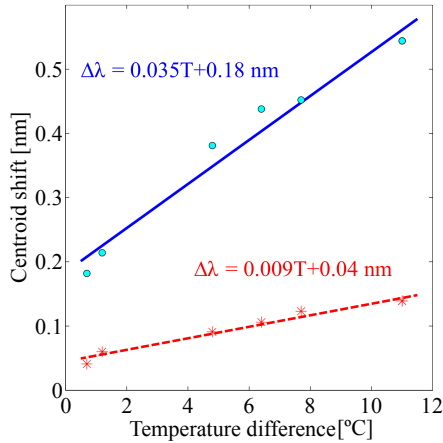}
\caption{Variation of the centroid of the output spectra as a
function of the temperature difference for a fixed post-selection
angle. The case $\beta=0^{\circ}$ (asterisks) illustrates the
situation where no WVA scheme is used. The case
$\beta=-40^{\circ}$ (circles) shows the case where the centroid
shift is enhanced by a factor of $\approx 4$ by means of the the
WVA scheme.}\label{fig:figure4}
\end{figure}

Figure \ref{fig:figure2} depicts the spectra of the signals
reflected from $\mathrm{FBG_1}$ ($\beta=-90^{\circ}$) and
$\mathrm{FBG_2}$ ($\beta=0^{\circ}$). Figure \ref{fig:figure2} (a)
corresponds to the signal reflected from $\mathrm{FBG_2}$ at a
fixed temperature, while Fig. \ref{fig:figure2} (b) corresponds to
the reflection from $\mathrm{FBG_1}$ at different temperatures.
The spectrum of the signal reflected from each FBG is composed of
a principal lobe $\sim 2.5\,\mathrm{nm}$ wide (FWHM), and a side
lobe with smaller amplitude that appear as a result of the high
contrast in index of refraction in the gratings. Since our scheme
relies on the measurement of the centroid of a Gaussian-like
spectra given a post-selection, the presence of non-negligible
side lobes can alter the measurements. To avoid this effect, each
measured spectrum is filtered numerically using a super-Gaussian
filter indicated by the dashed line in Fig. \ref{fig:figure2}.

Figure \ref{fig:figure3} (a) presents the measured shift of the
centroid of the output spectrum for $T_1-T_2=11^{\circ}\mathrm{C}$
and different post-selection angles (dots). For small angle
deviations around $\beta=-45^{\circ}$, shifts of the centroid of
the spectrum up to $\pm 0.6\,\mathrm{nm}$ are observed. This
corresponds to a three-fold enhancement with respect to the initial
shift of $0.19\,\mathrm{nm}$ given by the FBGs with no weak
amplification scheme. The solid line indicates the best
theoretical fit obtained using Eq. (\ref{deltanu_ref}). Fig.
\ref{fig:figure3} (b) shows some selected spectra measured after
performing the super-Gaussian filtering. In general, there is a
trade-off between the centroid shift observable for a specific
temperature difference and the amount of losses that can be
tolerated to keep a good SNR. In our scheme the side lobes become
relevant with respect to the main lobe of the output spectrum for
post-selection angles within the interval
$-50^{\circ}\leq\beta\leq -40^{\circ}$. For this reason, a maximum
amplification factor of $\approx 4$ is obtained for such angles.

Figure \ref{fig:figure4} shows the measured variation of the
centroid position of the output spectra as a function of the
temperature difference for a given post-selection. Circles
indicate the case when the output signal is projected into a
polarization state with $\beta=-40^{\circ}$, so that the output
spectrum centroid drifts $\sim 0.035\mathrm{nm/^{\circ}C}$ (solid
line). For the sake of comparison, asterisks show the case where
no weak amplification is used ($\beta=0^{\circ}$), generating a
spectrum centroid variation of $\sim 0.009\mathrm{nm/^{\circ}C}$
(dashed line). The use of the WVA provides a four-fold enhancement
of the sensitivity.

In conclusion, we have demonstrated that WVA can be used to
enhance the sensitivity of sensors based on  Fiber Bragg Gratings.
The shift of the spectrum of the signal reflected by a FBG due to
temperature changes was measured to be $\sim
0.009\mathrm{nm/^{\circ}C}$. With a weak amplification scheme, we
measured a  change of $0.035\mathrm{nm/^{\circ}C}$, a fourfold
increase. In scenarios where the measurable shift of the spectrum
is limited by the detection stage, but the decrease of signal
energy that accompanies still keeps the signal-to-noise ratio at
an usable level, weak value amplification is a promising scheme to
enhance the capabilities of FBG-based sensor systems.


\begin{thebibliography}{99}
\bibitem{book_kashyap} R. Kashyap, {\em Fiber Bragg Gratings}, Academic
Press, 1999.

\bibitem{aharonov1988} Y. Aharonov, D. Z. Albert, and L. Vaidman, Phys. Rev. Lett. \textbf{60}, 1351 (1988).

\bibitem{duck1989} I. M. Duck, P. M. Stevenson, and E. C. G. Sudarhshan, Phys. Rev. D \textbf{40}, 2112 (1989).

\bibitem{howell2010} J. C. Howell, D. J. Starling, P. B. Dixon, K. P. Vudyasetu, and A. N. Jordan, Phys. Rev. A \textbf{81}, 033813 (2010).

\bibitem{high_signal2012} J. P. Torres, G. Puentes, N. Hermosa, and L. J. Salazar-Serrano, Opt. Express \textbf{17} 18869 (2012).

\bibitem{ritchie1991} N. W. Ritchie, J. G. Story, and R. G. Hulet, Phys. Rev. Lett. \textbf{66} 1107-1110 (1991).

\bibitem{hosten2008} O. Hosten and P. Kwiat, Science \textbf{319}, 787 (2008).

\bibitem{ben_dixon2009} P. Ben Dixon, D. J. Starling, A. N.
Jordan, and J. C. Howell, Phys. Rev. Lett. \textbf{102}, 173601
(2009).

\bibitem{howell2010_freq} J. C. Howell, D. J. Starling, P. B. Dixon, K. P. Vudyasetu, and A. N. Jordan, Phys. Rev. A \textbf{82}, 063822 (2010).

\bibitem{xu_guo2013} X.-Y. Xu, Y. Kedem, K. Sun, L. Vaidman, C.-F. Li and G.-C. Guo, Phys. Rev. Lett. \textbf{111}, 033604 (2013).

\bibitem{salazar2014} L. J. Salazar-Serrano, D. Janner, N. Brunner, V. Pruneri, J. P. Torres, Phys. Rev. A \textbf{89} 012126 (2014).

\bibitem{tahir2009} B. A. Tahir, J. Ali, and R. A. Rahman, Int. J. Mod. Phys. B \textbf{23}, 2349 (2009).

\bibitem{ricchiuti2014} A. L. Ricchiuti, D. Barrera, K. Nonaka and
S. Sales, Opt. Lett. \textbf{39} 5729 (2014).

\bibitem{egan2012} P. Egan and J. A. Stone, Opt. Lett. \textbf{37}, 4991 (2012).

\bibitem{salazar_interference} L. J. Salazar-Serrano,
A. Valencia, and J. P. Torres, Opt. Lett. \textbf{39}, 4478
(2014).
\end{thebibliography}
\end{document}